\documentstyle[twocolumn,aps,epsf]{revtex}
\newcommand     {\jou}[4]    {{#1}{\bf{#2}}, {#3} ({#4}). }
\newcommand     {\PRL}   {Phys.Rev.Lett.\ }
\begin{document}
\thispagestyle{empty}
\noindent
{\bf Comment on "Finite Size Scaling in Neural Networks"} \\ \ \\

Nadler and Fink use finite size scaling (FSS) in [1] to estimate
the storage capacity of Ising neural networks from the behavior
for small $N$ of $P(\alpha,N)$, the probability that $\alpha N$
patterns can be loaded onto a network of size $N$.
For the perceptron they obtain a capacity
of $\alpha_c = 0.796\pm 0.010$, implying that the analytical result
$\alpha_c \approx 0.833$
(Ref. [26] of the letter) is wrong. This is remarkable since it is
hard to see how the latter result might be corrected within
the replica theoretic framework used in its derivation. Our
comment, however, demonstrates that FSS cannot be used to extrapolate
to large $N$ from the small system sizes considered by Nadler and
Fink. Consequently their work does not provide evidence that
the analytical result is incorrect.

By definition $P(\alpha,N)$ is monotonic in $\alpha$. If $\alpha$
and $N$ are such that the equivalence between the two quantities
established by FSS holds, $P(\alpha,N)$ must
be well approximated by a function monotic in $N$ for fixed
$\alpha$. The nonmonotonic
behavior of $P(\alpha,N)$ is shown in Fig. 1 for $\alpha_1 = 0.8$.
Thus it is methodologically flawed, to apply
FSS in the range of small $N$ $(N \le 30)$ considered in the letter.
Even for $\alpha = 0.810$, we have found that $P(\alpha,N)$ increases
with the system size (Fig. 1). Reasonably this can be taken as evidence
that  $\alpha_c > 0.810$, a finding which is hardly compatible with
the estimate of Nadler and Fink.  It is however difficult to see,
how a precise extrapolation to the infinite system might be obtained
from such small values of $N$.

We have been able to consider somewhat larger systems than in
the letter since we do not use exhaustive search. Instead we exploit the
fact that the loading problem for the perceptron may be efficiently
decided by the simplex algorithm in the case of continuous couplings:
Consider a binary search tree of depth $N+1$ and identify a node of depth
$M+1$ with an assignment of $1$ or $-1$ to the first $M$ couplings
$(J_1,\ldots,J_M)$. At such a node one may ask whether continuous couplings
$J_{M+1},\ldots,J_N$ in the interval $[-1,1]$ exist such the the
entire perceptron $(J_1,\ldots,J_N)$ stores a given training set. If
not, the subtree of the node may be discarded. Traversing the
search tree but pruning subtrees in the described manner yields
an algorithm which decides the loading problem for binary couplings.
For values of $\alpha$ around $0.8$ we have found the time complexity
of this algorithm to scale roughly as $2^{N/3}$, much better than
exhaustive search. This has enabled us to estimate $P(\alpha,N)$
rather accurately by conducting  at least $3\times 10^4$ trials for each
data point. As in the letter, training sets with Gaussian inputs
where used.

\ \\

\noindent
M. Schr\"oder and R. Urbanczik\\
Institut f{\"u}r Theoretische Physik\\
Universit{\"a}t W{\"u}rzburg, D-97074 W{\"u}rzburg\\
\\
Received:\\
PACS numbers: 87.10.+e, 02.70.Lq, 05.50.+q, 64.40.Cn

\begin{figure}
\hbox{
\epsfysize 8cm
\hbox to \hsize{\hss
\epsfbox{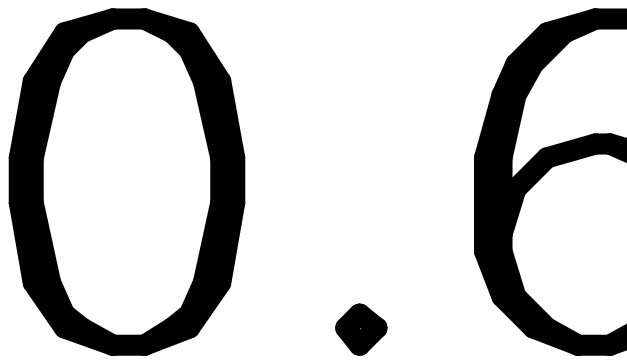} \hss}}
\caption{$P(\alpha,N)$ as a function of $N$ for $\alpha=4/5=0.8$ (upper graph)
and $\alpha=17/21\approx0.810$ (lower graph).
The values for the lower curve are: $P(17/21,21)=0.6660\pm0.0007$ and
$P(17/21,42)=0.6696\pm0.0020$.
Standard errorbars are shown.
}

\end{figure}

\end{document}